\documentclass[12pt]{article}
\begin{document}
\input epsf
\title{Ultrahigh Energy Neutrinos and the Highest Energy Cosmic Rays}
\author{G.~Domokos and  S.~Kovesi-Domokos\\
Department of Physics and Astronomy\\
The Johns Hopkins University\\
Baltimore, MD 21218\thanks{e-mail: skd@jhu.edu}\and Paul~T.~Mikulski\\
Department of Physics\\ United States Naval Academy\\
Annapolis, MD 21402\thanks{e-mail: mikulski@brass.mathsci.usna.edu}}
\date{June 2000}
\maketitle
\begin{abstract}
It has been suggested that the characteristic energy of string models may be 
con\-sid\-er\-ably lower than the observed Planck mass \mbox{($\approx 10^{19}$GeV)}. In  such
schemes, the unification of interactions takes place around the string scale, perhaps
as low as a few tens of TeV. Consequently, at energies above the string scale, 
neutrinos acquire interactions comparable in  strength to strong interactions. 
While they can propagate through the CMBR essentially uninhibited, in interactions
with nuclei in the atmosphere they induce air showers comparable to proton induced ones.
We conjecture that air showers above the Greisen-Zatsepin-Kuzmin (GZK) cutoff in the
cosmic radiation are induced by such neutrinos. A Monte Carlo simulation shows that
neutrino induced ``anomalous'' showers are virtually indistinguishable from proton induced ones
on an event-by event basis. However, given sufficient statistics in detectors 
(HiRes, OWL, Auger, $\ldots$), the post-GZK showers are expected to  exhibit characteristics 
 in the fluctuation pattern allowing a distinction
between proton and neutrino induced showers.
\end{abstract}
\vspace{2mm}
\begin{flushleft}
Paper submitted to Neutrino2000, Sudbury June 2000\\
\end{flushleft}
\section{Introduction}
The propagation of the highest energy cosmic rays (assumed to be protons) is
limited predominanly by pion photoproduction on the photons of the cosmic microwave 
background (CMBR).
This is the well known Greisen, Zatsepin, Kuzmin (GZK) effect, leading to a cutoff 
in the spectrum of primary cosmic rays at around
$6\times 10^{19}$eV in energy. No  source of high energy protons can be much farther 
 than about 20Mpc if the protons are to reach us without a substantial energy loss.
A modern and careful 
calculation
of the effect has been carried out by Hill and Schramm, ref.~\cite{hillschramm}which also
contains references to the original papers. The physics of the GZK effect is very well known
and it is not controversial: in fact, the  energy in the CM system of the collision 
between a cosmic ray proton and a typical photon of the CMBR is just about sufficient to
excite the $\Delta$ resonance. Hence, one is dealing with low energy hadron physics
explored for the past 45 years or so. 

As a consequence, the observation of primary cosmic rays well above the GZK cutoff
is a  puzzle. For a sampling of the observations, one can consult
a number of references, such as \cite{takeda} (AGASA) , \cite{bird} (Fly's~Eye)and  
Szabelski's  review,~\cite{szabelski}. In addition, the 
home page of
the AGASA detector,~\cite{agasa} contains  frequently  updated information
on the highest energy events observed. Apparently, there are no astronomical
objects within 50~Mpc or so from the Milky Way capable of producing particles
of the order of $10^{20}$eV, with the possible exception of M87, 
{\em cf\/} Biermann~and~ Strittmatter,~ref.~\cite{biermann}.

It is to be noted that
the highest energy event observed by Fly's~Eye, see ref.~\cite{bird} appears to
have generated an extensive air shower (EAS) closely resembling one
generated by a proton, as shown in ref.~\cite{delaware}. However, due to fluctuations in the 
development of an EAS,
{\em one} event alone cannot uniquely determine the nature of the primary
particle. A satisfactory resolution of this question requires a
substantial amount of data collected by present and future detectors, such as
HiRes, OWL, AIRWATCH, Auger {\em etc.\/} 

Assuming the puzzle to be a real one, there are basically two types of explanations 
to be found in the literature.
\begin{itemize}
\item {\em Astrophysical ones}, with the work in ref.~\cite{ahn} being the most recent 
(and most credible) one. The authors of that reference assume that most (all?) of 
the post-GZK events are protons originating from M87. The observed near-isotropy of the
distribution is explained by postulating a galactic wind.
\item {\em Physics beyond the Standard Model or rare processes within the
framework of the SM\/.}
A  fair sampling of those  is
contained in the proceedings of the University of Maryland {\em Workshop
on Observing Giant Cosmic Ray Air Showers}~\cite{krizmanic}.
\end{itemize} 

In the light of 
recent, accelerator based experiments at LEP and the Tevatron, the only proposal based on 
the SM and its supersymmetric extensions which remains plausible  is 
Weiler's~\cite{weiler}. In essence, Weiler proposes that UHE energy neutrinos 
interact with relic ones in our ``cosmic neighborhood'' and excite the $Z$ resonance.
The $Z$, in turn, decays predominatly into quark pairs. Hence, a proton can be created
sufficiently close to us in order to evade the GZK cutoff.

Our proposal~\cite{prl99}, following up on an earlier one~\cite{domonuss},
similarly conjectures that the post--GZK events are 
caused by neutrinos. Both Weiler and we agree that neutrinos penetrate the CMBR essentially 
uninhibited: the typical $\sqrt{s}$ in an interaction between an UHE neutrino and a photon
of the CMBR is of the order of $100$MeV. This is in the realm of the SM and, in essence,
the UHE neutrino does not interact with the CMBR.

In contrast to Weiler, however, we conjecture that the post--GZK events originate in the 
atmosphere, due to new physics. This may have some advantages as far as the energetics at the
source  of neutrinos is concerned.  Moreover, as soon as a sufficient number of post--GZK
events will be collected, the hypothesis will become relatively easily testable.

The approximate isotropy of the post-GZK events receives the same explanation in 
Weiler's scenario as in  ours: neutrinos do not interact with the CMBR and UHE 
neutrinos generated by a multitude of sources reach us uninhibited.

In the following section we briefly outline the argument leading to a {\em
precocious unification and some of its consequences}, based on ref.~\cite{prl99}. New results
are presented in the section describing the MC simulation of post-GZK showers.
The last section contains a discussion of the results.
\section{Precocious Unification}
``Old fashioned'' grand unification theories (GUT) as well as string models were based on the
notion that the unification of forces (including gravity) can take place only around
the Planck energy. Recent work by Lykken~\cite{lykken}, Dimopoulos~{\em et al.\/}~\cite{dimo}, 
Dienes~{\em et al\/.}~\cite{dienes}\footnote{Due to the rapidly increasing number of works 
on this subject, here we can only cite
the {\em earliest} articles of these authors on the topic.}questions this dogma, by 
pointing out that the existence
of extra (probably compactified) dimensions in various string models allows one to 
separate the string scale from the observed Planck scale \mbox{($M_{P}\simeq 10^{19}$GeV).}
In fact, the string scale can be as low as a few or a few tens of a TeV, without violating known
experimental constraints, including the lifetime of the proton.

It has to be emphasized that such a scenario lacks, at this time, a solid dynamical underpinning.
Nevertheless, it is very interesting from the experimental/observational point of view, and,
most importantly, its main consequences can be tested within the next decade.

As it was pointed out in ref.~\cite{prl99}, low mass scale {\em string}-based unification 
implies a rapidly ({\em exponentially}) rising level density of intermediate  excited 
states involved in any  given
reaction at energies either soon to be available for experimentation (LHC) or at modern cosmic 
ray detectors\footnote{One recalls that if a particle  -- of almost any species -- at a 
laboratory 
energy of the order of $10^{20}$eV interacts with a nucleus in the atmosphere, the 
CMS energy of the reaction is of the order of a few hundred TeV.}.
As a consequence, cross sections of essentially {\em all} relevant reactions
reach their value dictated by the unified theory very rapidly. This is a pleasing and almost
model independent consequence of such scenarios: all string models give rise to an
 exponentially rising level density of excited states\footnote{We thank K.~Dienes for a
correspondence on this topic.}.
(The transition from a logarithmical to a power behavior of the running couplings has 
been particularly stressed  by Dienes~{\em et al\/.}~\cite{dienes}). 
In ref.~\cite{prl99}, we could merely test the plausibility of the scenario outlined above.
It was found that with the string entropy growing as $S\simeq \sqrt{N}$ and with 
reasonable structure functions, a cross section of the order of the strong one can be reached
at laboratory energies ranging between $10^{19}$ and $10^{20}$ eV. 
The characteristic energy (inverse Regge slope) required for this  is of the order of a few TeV.
(Here $N$ stands for the index of the level of excitation.) In any  string model, a string
entropy given above translates into a level density rising as
\[  \rho \simeq \exp (s/s_{0})^{1/2}, \]
where $s_{0}$ is the inverse Regge slope. In most models there is a power behaved
prefactor in the expression of the level density. We found, however that the
results are insensitive to the prefactor and it was omitted.
\section{Characteristics of the ``anomalous'' showers.}
For the sake of simplicity, neutrino induced showers in the energy region around and
above the characteristic energy are henceforth called ``anomalous''. 

We assumed the following characteristics of the elementary processes giving rise to
anomalous showers.
\begin{itemize} 
\item Around the CMS energy $\simeq \sqrt{s_{0}}$, the neutrino -- quark cross section 
begins to rise above its Standard Model value  
as dictated by the level density of $s$-channel excitations given above. The exponential 
rise continues until the cross section reaches a prescribed fraction (say, 1/2) of the
strong cross section. Thereafter, the cross section levels off: unitarity does not allow
cross sections which rise exponentially forever.
\item As long as $s$ remains larger than about $s_{0}$, 
any interaction produces quarks and leptons in roughly equal numbers. Once $s$
drops below $s_{0}$, the particles interact with cross sections as given by the
Standard Model. Quark production in lepton induced reactions and lepton
production in quark ({\em i.e.\/} hadron) induced reactions was neglected in the 
latter energy range.
\end{itemize}
By experimenting with a variety of functions describing the rise and leveling off of the
cross sections, we found that the final results were insensitive to the precise form
of the function. For that reason, most of the shower simulations were carried out
using a step function\footnote{It was pointed out by Burdman {\em et al.\/} \cite{burdman} 
that, strictly speaking, any step function threshold violates unitarity. This occurs because if
the imaginary part of an amplitude has a step function discontinuity, its real part
is (logarithmically) infinite at the point of discontinuity. However, as long as we deal 
with cross sections only, approximating a rapid rise by a step function does no harm.}.

The development of the ``anomalous'' showers was modeled by means of a one dimensional
MC. Standard model processes were modeled along fairly standard lines. One of the main
innovations in the program was that input data  could be modeled in a flexible way, so that it
was relatively easy to experiment with various assumptions. The model is described 
in detail in ref.~\cite{pault}. 

Here we present data assuming that at unification the cross section is approximately
1/2 of a SM hadronic cross section, extrapolated to the characteristic energy ($s_{0}$)
by means of a quadratic polynomial in $\ln s$.
In the following Figure we display the average longitudinal profile
of ``anomalous'' showers. For comparison, we also plot  the average 
longitudinal profile of a proton induced shower. The profile of the
proton induced shower is in reasonably good agreement with other calculations.
(It has to be noted that there exist considerable uncertainties in
a MC simulation of showers, largely due to the lack of direct
measurements of cross sections, multiplicities, {\em etc\/.} in the 
relevant energy region. For a detailed discussion, {\em cf\/.} 
\cite{pault}.)

\epsfxsize=\textwidth \epsfbox{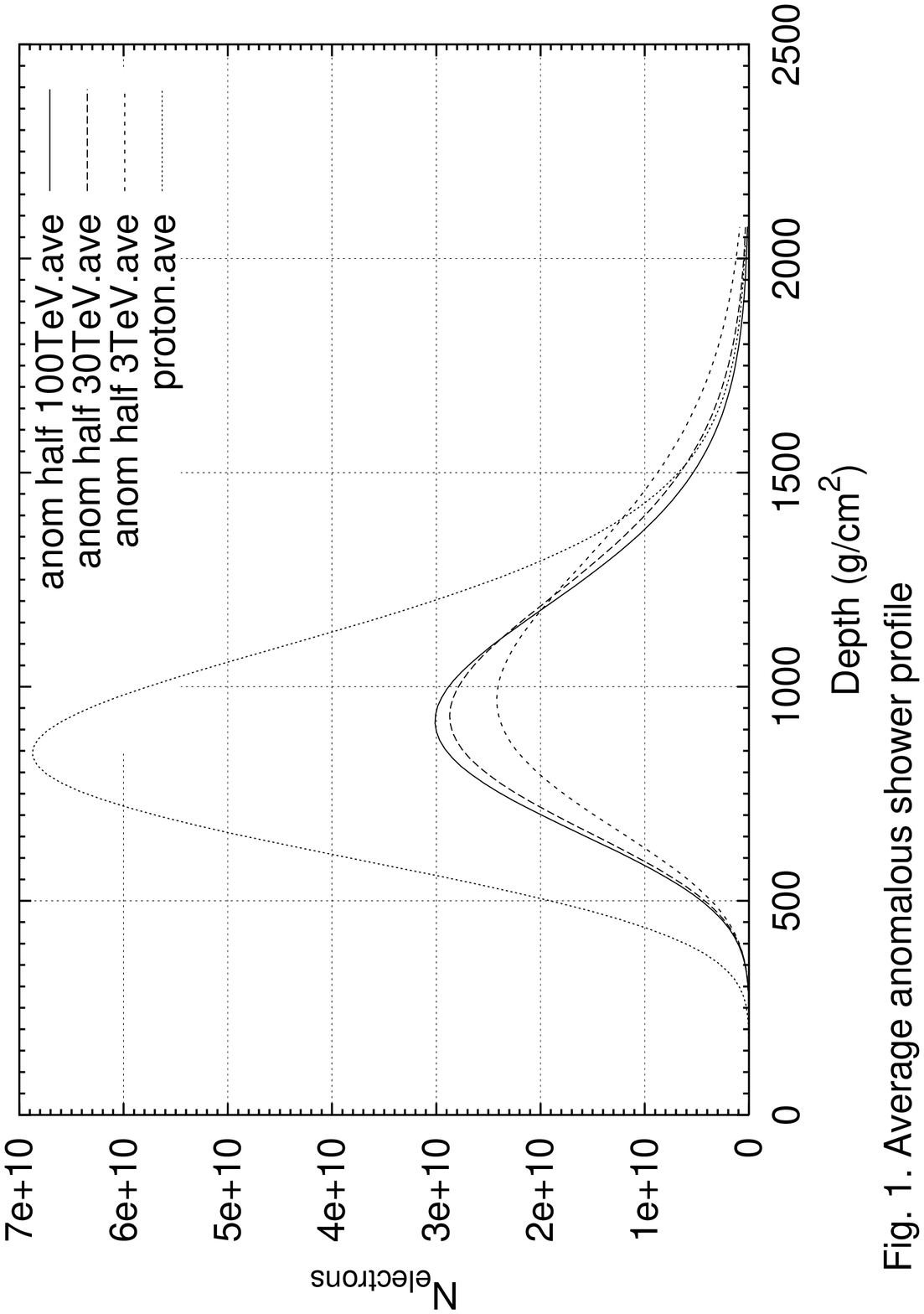}

The development of the ``anomalous'' showers has been simulated for three
values of $s_{0}$, as shown in Fig.~1. One sees a few prominent features in this Figure.
\begin{itemize}
\item The multiplicity of electrons around $\left< X_{max}\right>$ is about half of 
that contained in a 
proton induced shower. This is due to the fact that, if forces are unified, a 
substantial part of the primary energy goes into prompt lepton production; lepton 
interaction cross sections and multiplicities are lower than hadronic ones.
Consistent with this picture is the result that the electron deficiency {\em increases}
with {\em decreasing} $s_{0}$. For a lower characteristic energy, the prompt lepton
production due to unification takes place for a longer portion of the shower after the 
first interaction.
\item For the same reason as stated above, the position of $\left< X_{max}\right> $ is 
somewhat deeper 
than in a proton induced shower. However, the value  of $\left< X_{max}\right> $ is, 
apparently, a
rather slowly varying function of $s_{0}$.
\end{itemize}

It is not very likely that such features can be distinguished on an {\em event-by event} basis.
For instance, if the electron number is smaller, one is likely to interpret the event as
having a smaller primary energy. Likewise, an $X_{max}$ larger than the expected one
(of the order of $850 {\rm g/cm^{2}}$) is likely to be interpreted as a fluctuation
in the shower development.

There is, however, a {\em substantial} difference in the fluctuations  around
the shower maximum. In view of the fact that in the near future one is likely to have only a
limited number of post-GZK events, we decided to characterize the fluctuations by a single 
parameter, namely the rms deviation from the mean value of $X_{max}$. In Figure~2,
we plotted the rms deviations for the same values of $s_{0}$ as in Fig.~1 and again,
for comparison, the rms fluctuation for proton induced showers.

\epsfxsize=\textwidth \epsfbox{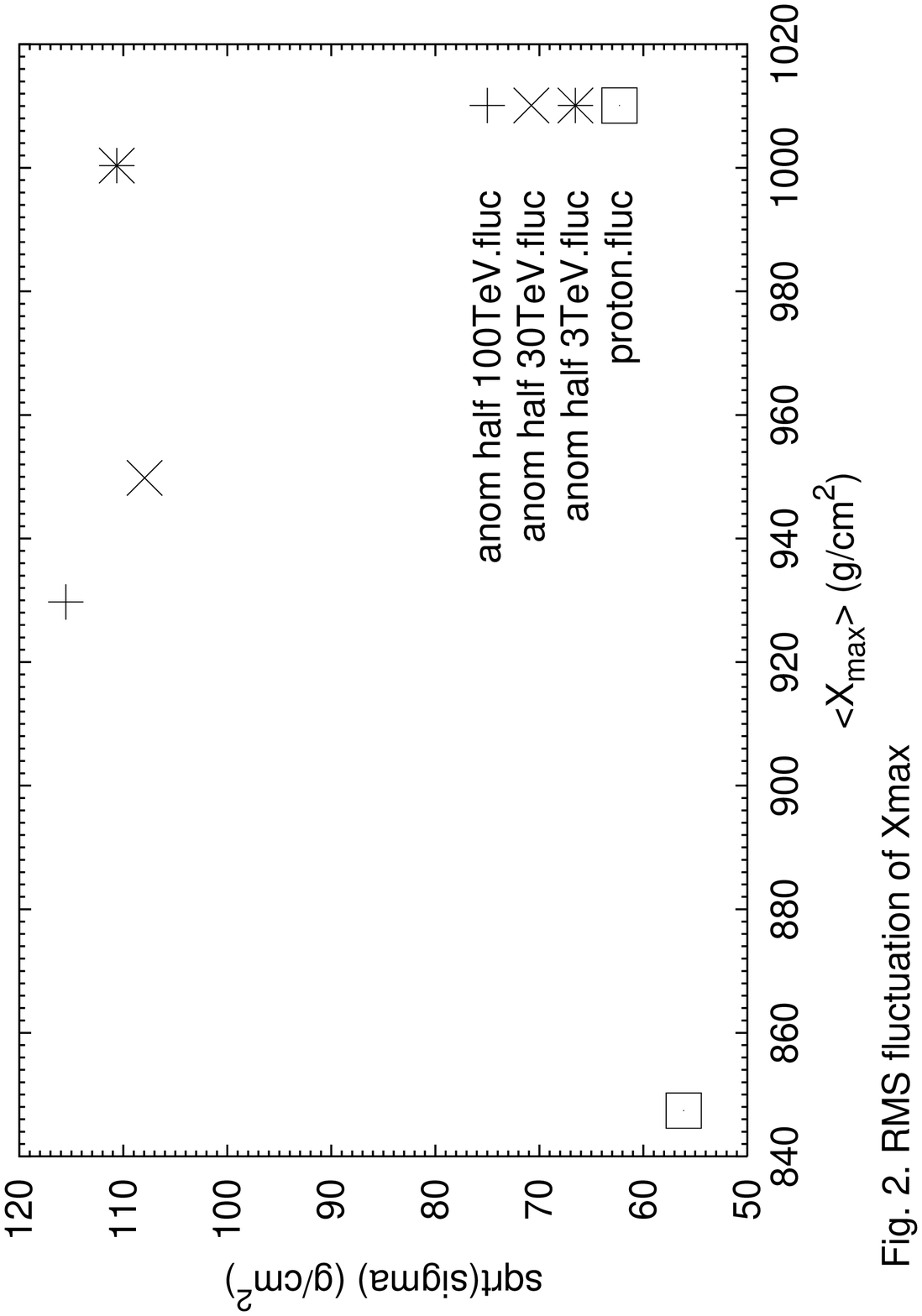}

There are two main features discernible in Fig.~2.
\begin{itemize}
\item The rms fluctuation around $\left< X_{max}\right>$ increases from about $55 {\rm g/cm^{2}}$
(proton induced showers) to a fluctuation about a factor of 2 or so larger in the case of
``anomalous'' showers. 
\item As in Fig.~1, the dependence on $s_{0}$ is weak. In this MC 
simulation, we suspect that the differences between rms fluctuations for various values
of $s_{0}$ are largely due to statistical fluctuations in the simulation itself: there 
appears to be no systematic trend in the correlation between $\left< X_{max}\right>$
and the rms fluctuation.
\end{itemize}

It is easy to understand the main features displayed in Fig.~2 in terms of the
``new physics'' involved.  It is well known that in the development of 
a cascade, if the latter is dominated by processes of small cross section and/or small
multiplicities, the cascade exhibits large fluctuations. The ``new physics'' 
as conjectured here, contributes in this way to the initial stages of the cascade;
hence, a somewhat dramatic increase of the fluctuations comes as no surprise.
\section{Discussion}
Even though there is no reliable dynamical theory describing low scale string physics yet,
the basic aspects of the scenario exploited here are  very attractive. (Among other
things, the hierarchy problem of interactions and masses of elementary particles is
likely to be alleviated. It was also pointed out by Dienes {\em et al\/.}~\cite{dienes} 
some time
ago that a low scale unification does not have to lead to a rapid proton decay
as it was previously believed.) 

The study of the highest energy cosmic rays probably provides an interesting laboratory 
for the study of these ideas, complementing lower energy, accelerator based experiments.

Some comments are in order regarding the results presented here and on open questions.
As emphasized in the preceding Section, all our results depend rather weakly on the
magnitude of the characteristic energy. Preliminary calculations also indicate
that variations of the cross section at unification does not affect the qualitative
aspects of the results very much. For instance, if we assume that the cross section at 
unification equals the extrapolated value of the hadronic cross section, 
$\left< X_{max}\right>$ gets somewhat closer to its value in proton induced showers.
Likewise, the rms fluctuation around $X_{max}$ becomes somewhat smaller. However,
the shower is not identical to a proton induced one. This is due to the fact that
in the first few interactions, roughly half of the energy ends up in leptons; when
the energy drops below its critical value, the latter contribute to the evolution of
the shower through low multiplicity interactions.

One of the important open questions is about the astrophysical origin
of UHE neutrinos. It appears that the production of particles of any kind
of energies around $10^{19}$eV and above is an unsolved and challenging
problem in astrophysics. If one wants
to remain within the framework of the SM, the production of neutrinos of similar 
energies can take place as a result of the weak decays of pions and other
hadrons. In order to excite the $Z$ resonance on relic neutrinos, the incident 
neutrino energy has to be  of the order of $10^{24}$eV.
 Using a straightforward extrapolation of known
hadronic cross sections and multiplicities, one concludes that protons of even higher energies
are needed.  This may seriously aggravate the 
astrophysical problem.

If, however, the incident neutrino initiates a  shower on an
``air nucleus''\footnote{An ``air nucleus'' is the average of  $O$ and $N$ nuclei,
weighted with the concentration of each in the atmosphere.}, it only needs an energy 
approximately equal to that of
an incident proton. It is not clear at present whether the ``new physics'' can
contribute to the solution of the astrophysical problem.

\end{document}